\documentclass[11pt,a4paper,pdftex]{article}
\pdfoutput=1
\usepackage{jcappub}
\usepackage{color}

\newcommand{\be}{\begin{eqnarray}}
\newcommand{\ee}{\end{eqnarray}}
\newcommand{\rar}{\rightarrow}

\title{Black hole evaporation in conformal gravity}

\author[a,b]{Cosimo~Bambi,}
\author[c,a]{Leonardo~Modesto,}
\author[d]{Shiladitya~Porey,}
\author[a]{Les\l{}aw~Rachwa\l{}}

\affiliation[a]{Center for Field Theory and Particle Physics and Department of Physics,\\
Fudan University, 220 Handan Road, 200433 Shanghai, China}
\affiliation[b]{Theoretical Astrophysics, Eberhard-Karls Universit\"at T\"ubingen,\\ 
Auf der Morgenstelle 10, 72076 T\"ubingen, Germany}
\affiliation[c]{Department of Physics, Southern University of Science and Technology,\\ 
1088 Xueyuan Road, Shenzhen 518055, China}
\affiliation[d]{Department of Physics, Indian Institute of Technology, 208016 Kanpur, India}

\emailAdd{bambi@fudan.edu.cn}
\emailAdd{lmodesto@sustc.edu.cn}
\emailAdd{shilp@iitk.ac.in}
\emailAdd{rachwal@fudan.edu.cn}

\abstract{We study the formation and the evaporation of a spherically symmetric black hole in conformal gravity. From the collapse of a spherically symmetric thin shell of radiation, we find a singularity-free non-rotating black hole. This black hole has the same Hawking temperature as a Schwarzschild black hole with the same mass, and it completely evaporates either in a finite or in an infinite time, depending on the ensemble. We consider the analysis both in the canonical and in the micro-canonical statistical ensembles. Last, we discuss the corresponding Penrose diagram of this physical process.}


\begin{document}

\maketitle

%%%%%%%%%%%%%%%%%%%%%%%%%%%%%%%

\section{Introduction}

Einstein's theory of gravity is currently the standard framework for the description of classical gravitational fields and of the geometrical structure of the spacetime. It has successfully passed a large number of observational tests~\cite{will}. Nevertheless, it is plagued by a number of conceptual problems that show up in some extreme conditions. One of them is the appearance of spacetime singularities, where predictability is lost and standard physics breaks down.

In the past decades, there have been significant efforts to try to resolve these singularities by employing different generalizations of the theory. One of these possibilities is represented by the family of conformal theories of gravity~\cite{cg1,cg2,cg3,cg4,cg5,cg6,cg7,cg8,cg9}. In this context, the theory is invariant under a conformal transformation of the metric tensor
\be
g_{\mu\nu} \rar \Omega^2 g_{\mu\nu} \, ,
\ee
where $\Omega = \Omega(x)$ is a function of the spacetime point. Einstein's gravity is clearly not conformally invariant in a standard way, but the conformal transformations can be easily incorporated as a symmetry of the theory as we shall show below. A simple example of a conformally invariant theory is described by the action
\be  
S = - 2 \!\int\! d^4 x \, \sqrt{|g|} \left( \phi^2 R + 6 \, g^{\mu\nu} (\partial_\mu \phi)( \partial_\nu \phi) \right) .
\label{confEgrav}
\ee
The conformal compensator field $\phi$ (dilaton) transforms homogeneously: $\phi\to\Omega^{-1}\phi$. The equations of motion of this theory are
\be\label{eq-1}
\phi^2 G_{\mu\nu} &=& \nabla_\mu \partial_\nu \phi^2 - g_{\mu\nu} \Box \phi^2 
 - 6 \left( (\partial_\mu \phi)( \partial_\nu \phi) 
- \frac{1}{2} g_{\mu\nu} g^{\alpha\beta} (\partial_\alpha \phi)( \partial_\beta \phi) \right)\!, \\
\label{eq-2}
\Box \phi &=& \frac{1}{6} R \phi \, .
\ee
Eqs.~(\ref{eq-1}) and (\ref{eq-2}) are conformally invariant. If $(g_{\mu\nu}, \phi)$ is a solution, then also $(g_{\mu\nu}^*, \phi^*)$ is a solution, where
\be
g_{\mu\nu}^* = \Omega^2 g_{\mu\nu} \, , \quad \phi^* = \Omega^{-1} \phi \, .
\ee
If $\phi = 1/\sqrt{32 \pi G_N}={\rm const}$, all the derivatives of $\phi$ vanish and we recover Einstein's gravity with the proper normalization.

Conformal gravity solves the problem of spacetime singularities present in Einstein's gravity. This is done by finding a suitable conformal transformation $\Omega=\Omega(x)$ and interpreting the resulting $g_{\mu\nu}^*$ as the ``physical'' metric of the spacetime. In particular, in Refs.~\cite{p1,noi} we found a singularity-free black hole solution. The line element is
\be\label{eq-k}
ds^2 &=& \left(1 + \frac{L^2}{r^2}\right)^4 ds^2_{\rm S} \, ,
\ee
where $L$ is a new length scale and $ds^2_{\rm S}$ is the line element of the Schwarzschild metric. Here the conformal factor $\Omega$ is
\be\label{OmegaOk}
\Omega = \left(1 + \frac{L^2}{r^2}\right)^2 \, .
\ee

The key-points of the metric in Eq.~(\ref{eq-k}) are that: $i)$ the curvature invariants do not diverge at $r=0$, and $ii)$ the spacetime is \emph{not} geodesically incomplete at $r=0$~\cite{p1,noi}. The line element in Eq.~(\ref{eq-k}) describes thus a singularity-free non-rotating black hole. Even after the conformal transformation, this is still a black hole solution, because the surface at $r_{\rm H}=2 G_N M$ is null and the interior is causally disconnected to future null infinity. The point $ii)$ is possible because the affine parameter of geodesics diverges at $r=0$, namely particles (both massless and massive) never reach the center of the black hole. In the case of a massive probe, the affine parameter of the geodesic can be identified with the proper time of the particle. Current astrophysical data can already constrain the spacetime metric around astrophysical black holes (for a review, see e.g.~\cite{rev1,rev2,lingyao} and references therein) and current bounds on $L$ are discussed in Ref.~\cite{noi2}. Other references on the theory can be found in~\cite{exactsol,super-ren0, super-ren1, super-ren2, sren,scattering,fingauge,entropy,cosmology}.

It is important to understand that there is a symmetric and a broken phase of the theory~\cite{noi}. In the symmetric phase, the theory (\ref{confEgrav}) is explicitly invariant under conformal transformations and all observable quantities are independent of the choice of the conformal 
factor $\Omega$. The latter is associated with the choice of the gauge and has no physical implications. However, the Universe around us is not conformally invariant: if we believe that conformal invariance is a fundamental symmetry of Nature, it must be somehow broken. One possibility is that conformal invariance is spontaneously broken. Nature must select one of the vacua, namely one of the conformal factors around which we define the perturbative quantum field theory. Physical quantities are different in the symmetric and in the broken phases. In the symmetric phase, we can not perform any measurement of lengths and time intervals because we can not define a standard rod and a standard clock. In the broken phase, different vacua can have different physical properties because they represent different configurations: the conformal symmetry is broken and the particular choice of the conformal factor produces observable effects. At the moment, we do not have a mechanism to select one particular  vacuum in the broken phase, although the metric must be in the class of singularity free spacetimes. Notice that we have potentially an infinite class of regular metrics~\cite{noi}. In this paper we consider the conformal factor~(\ref{OmegaOk}), but we could consider any other ``good'' conformal factor, namely any other conformal factor that makes the metric regular. However, there are some features common to all form factors, like the infinite amount of proper time to reach $r=0$, etc. All these comments are about the selection of the regular spacetime in a large class of singularity free spaces, but once one regular metric is selected out the conformal symmetry is explicitly broken by the solution itself because the metric depends on up two scales.

Notice that there is no gravitational constant $G_N$ in the action in the symmetric phase of the theory. Indeed, the Newton constant $G_N$ is replaced by the dilaton field in order to achieve conformal invariance. However, the metric in~(\ref{eq-k}) is an exact solution of the field equations of the theory and introduces two scales, namely $L$ and $r_{\rm H} = 2 G_N M$ and the conformal symmetry is explicitly broken by the solution as we already pointed out above.

The aim of this paper is to study the formation of a singularity-free black hole in conformal gravity and the subsequent evaporation due to Hawking effect. In Section~\ref{s-2}, we study the gravitational collapse of a spherically symmetric thin shell of radiation using the Vaidya metric ansatz and we find the creation of a singularity-free black hole. In Section~\ref{s-3}, we consider the thermodynamics of the conformally transformed Schwarzschild black hole and how it depends on the new scale $L$. In Section~\ref{s-4}, we discuss the Penrose diagram for the formation and evaporation of a black hole in conformal gravity. Summary and conclusions are in Section~\ref{s-5}. In what follows, we employ units in which $c = G_N = \hbar =k_B= 1$ and a metric with signature $(-++\,+)$.

\section{Gravitational collapse \label{s-2}}

Since massless particles are naturally conformally invariant, the simplest model of gravitational collapse in conformal gravity is represented by the collapse of a thin shell of radiation. We have the Vaidya metric~\cite{Vaidya:1951zz}
\be
ds^2_{\rm V} = - \left(1 - \frac{2 M(v)}{r}\right) dv^2 + 2 dv dr +  r^2 d \Omega^{(2)} \, , 
\label{Vaidya_metric}
\ee
where $d \Omega^{(2)} = d\theta^2 + \sin^2\theta d\phi^2$, and $M(v)$ is given by
\be
M(v) =   M_{0}\Theta (v-v_{0})\,,
\ee
where $v$ is the ingoing null coordinate. This metric is a solution in Einstein's gravity for a collapsing spherically symmetric shell of radiation. At the same time, this is also an exact  solution in conformally invariant Einstein's gravity~\eqref{confEgrav}~\cite{exactsol}. Here $M_{0}$ is the ADM mass and $\Theta$ is the unit step function (Heaviside function). The stress-energy tensor of the thin shell of radiation is
\be
T_{\mu\nu} = {\rm diag} \left( \frac{M_{0} \delta (v- v_{0})}{4 \pi  r^2}, 0, 0, 0 \right) 
\label{Tmunu} \, .
\ee
One can check that using the metric \eqref{Vaidya_metric} the above stress-energy tensor is traceless and this is an important consistency requirement when it is coupled to conformal gravity.

In conformal gravity, we apply the conformal transformation proposed in~\cite{p1,noi} and we find
\be
ds^2 = \left(1 + \frac{L^2}{r^2}\right)^4
ds^2_{\rm V} \, .
\ee
The scalar field $\phi$, which had the profile  $1/\sqrt{32 \pi}$ in Einstein's gravity, now becomes
\be
\phi = \frac{1}{\sqrt{32 \pi}} \left(1 + \frac{L^2}{r^2}\right)^{\!\!-2} \, .
\ee
The scalar invariants for this conformally transformed Vaidya metric are calculated in Appendix~\ref{s-a} and they are regular for $L \neq 0$ at all radii $r$ and all values of the  null ingoing coordinate $v$.

The interpretation of the conformal Vaidya spacetime is the same as in Einstein's gravity. The thin collapsing shell of radiation moves along the null geodesic $v = v_0$. Inside the shell, the spacetime is conformally flat. Outside the shell, the metric is described by the conformally modified static Schwarzschild metric as in Eq.~(\ref{eq-k}). However, in conformal gravity Birkhoff's theorem does not hold any more. The event horizon forms when the shell crosses the Schwarzschild radius $r_{\rm H} = 2 M$.  The difference with the solution in Einstein's gravity is only that now, thanks to conformal transformations, we have removed the central singularity of the metric. It is worth noting that the stress-energy tensor of the radiation in Eq.~\eqref{Tmunu} and the one of the scalar field $\phi$ diverge at $r\to0$, as well as their sum. However, these are not invariant quantities as they depend on the choice of the coordinate system. Due to the fact that all thermodynamical properties of the black hole depend on the situation on its horizon, we can analyze them by studying the static conformally transformed Schwarzschild solution. The latter exactly describes the spacetime outside the central region, where the metric is time-dependent because of the collapse of the shell.

\section{Black hole evaporation \label{s-3}}

Let us now study the evaporation process of the newly born black hole. For the sake of simplicity, here we consider the conformally modified Schwarzschild metric instead of the conformally modified Vaidya one. After the formation of the horizon, the two solutions describe the same physical system, and therefore it is an unnecessary complication to work with the conformally modified Vaidya metric. Moreover, we observe that in the late time regime the conformally modified Vaidya solution approaches the conformally modified static Schwarzschild metric.

The line element of a static and spherically symmetric black hole spacetime can always be written in the form
\be
ds^2 = g_{tt} dt^2 + g_{rr} dr^2 + r^2 d \Omega^{(2)} \, , 
\ee
where the metric coefficients $g_{tt}$ and $g_{rr}$ are independent of $t$, $\theta$, and $\phi$. With such a choice of the metric tensor, the surface gravity is~\cite{visser}
\be
\kappa =- \lim_{r \rar r_{\rm H}}  \frac{1}{2 \sqrt{|g_{tt} g_{rr}|}} \frac{\partial g_{tt}}{\partial r} \, ,
\ee
where $r_{\rm H}$ is the radial coordinate of the event horizon. The Hawking temperature of the black hole is simply
\be
T_{\rm H} = \frac{\kappa}{2 \pi} \, .
\ee
In what follows, it is also necessary to calculate the surface area at the event horizon
\be
A = \lim_{r \rar r_{\rm H}} \int_0^\pi \int_0^{2\pi} \sqrt{g_{\theta\theta} g_{\phi\phi}} \, d\theta \, d\phi \,.
\ee

If we employ the conformally modified Schwarzschild metric, then we find that the Hawking temperature of the black hole $T_{\rm H}$ is the same as in Einstein's gravity, while the surface area $A$ is different if $L \neq 0$:
\be
&& T_{\rm H} =\frac{1}{8 \pi M} \, , \\
&& A = \frac{\pi  \left(L^2+4 M^2\right)^4}{16 M^6} \, .
\ee
Since the relation between the Hawking temperature $T_{\rm H}$ and the black hole mass $M$ is the same in Einstein's and conformal gravity, the black hole entropy, as a function of the mass, does not change and it is given by the formula
\be
S(M)=4\pi M^2\,.
\label{entropy}
\ee
 The black hole luminosity (or mass loss rate) $L_{\rm H}$ due to Hawking radiation is instead different because the surface area now depends on $L$.

\subsection{Canonical ensemble}

The canonical statistical ensemble is defined by the condition that the energy (mass) is constant in the statistical equilibrium state. One finds for the black hole luminosity
\be\label{eq-dmdt}
L_{\rm H} = - \frac{dM}{dt} = \sigma A T^4_{\rm H} \, ,
\ee
where $\sigma$ is some constant that takes into account the particle content and also depends on the black hole mass~\cite{page1,page2}. The constant $\sigma$ would be the Stefan-Boltzmann constant ($\sigma_{\rm SB}= \pi^2 / 60$) if the black hole emission were exactly that of a black body. A black hole emits any particle of the theory (not only electromagnetic radiation) and the spectrum of the emitted particles deviates from the black body one due to the mass and the spin of the particles, the finite size of the black hole, etc. In the canonical ensemble, the system is completely stationary and the changes due to evaporation do not influence the energy of the state. This means that for all computations within the ensemble we can use the static conformally transformed Schwarzschild metric.

\begin{figure*}[t]
\vspace{0.3cm}
\begin{center}
\includegraphics[type=pdf,ext=.pdf,read=.pdf,width=7cm]{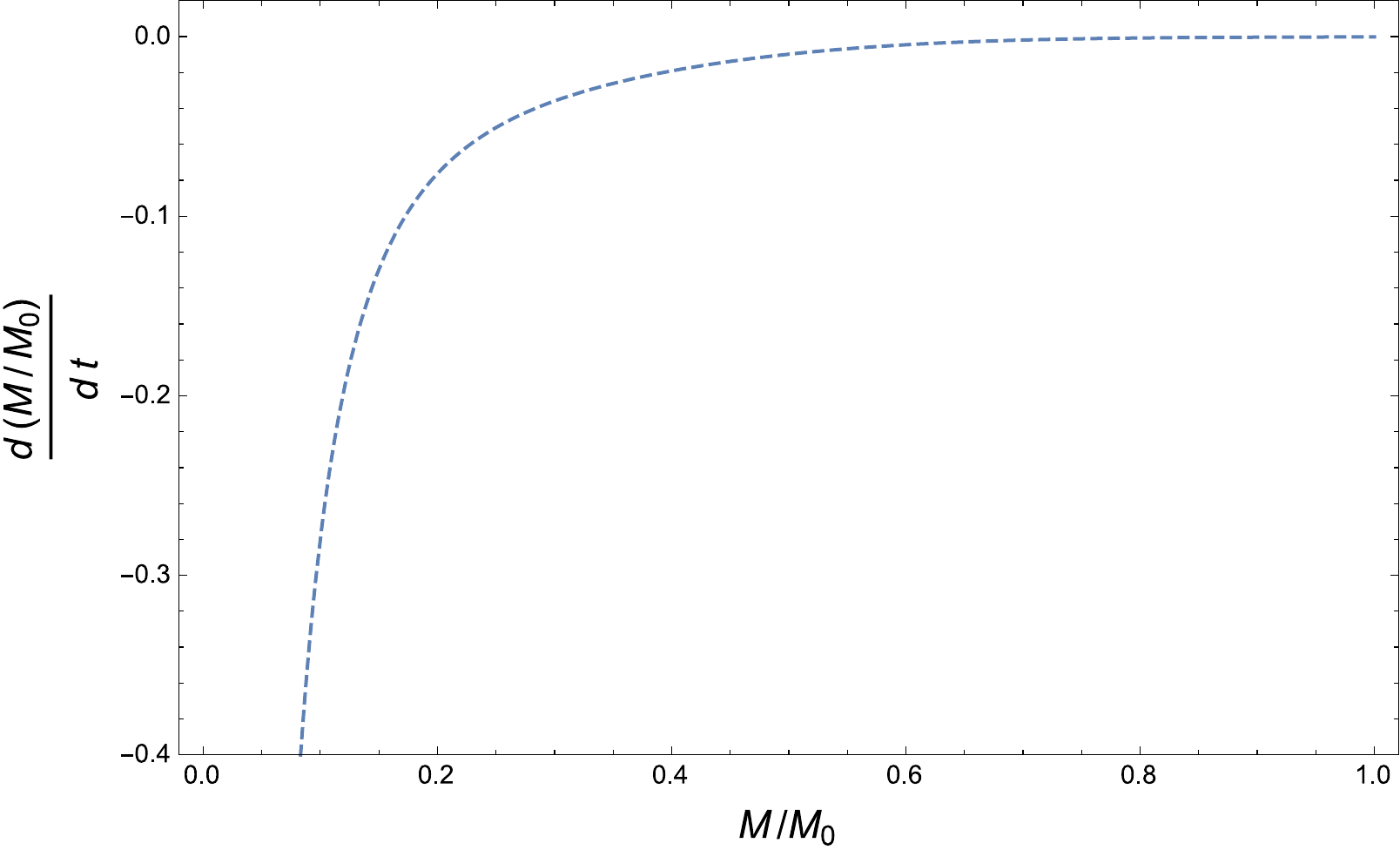}
\hspace{0.5cm}
\includegraphics[type=pdf,ext=.pdf,read=.pdf,width=7cm]{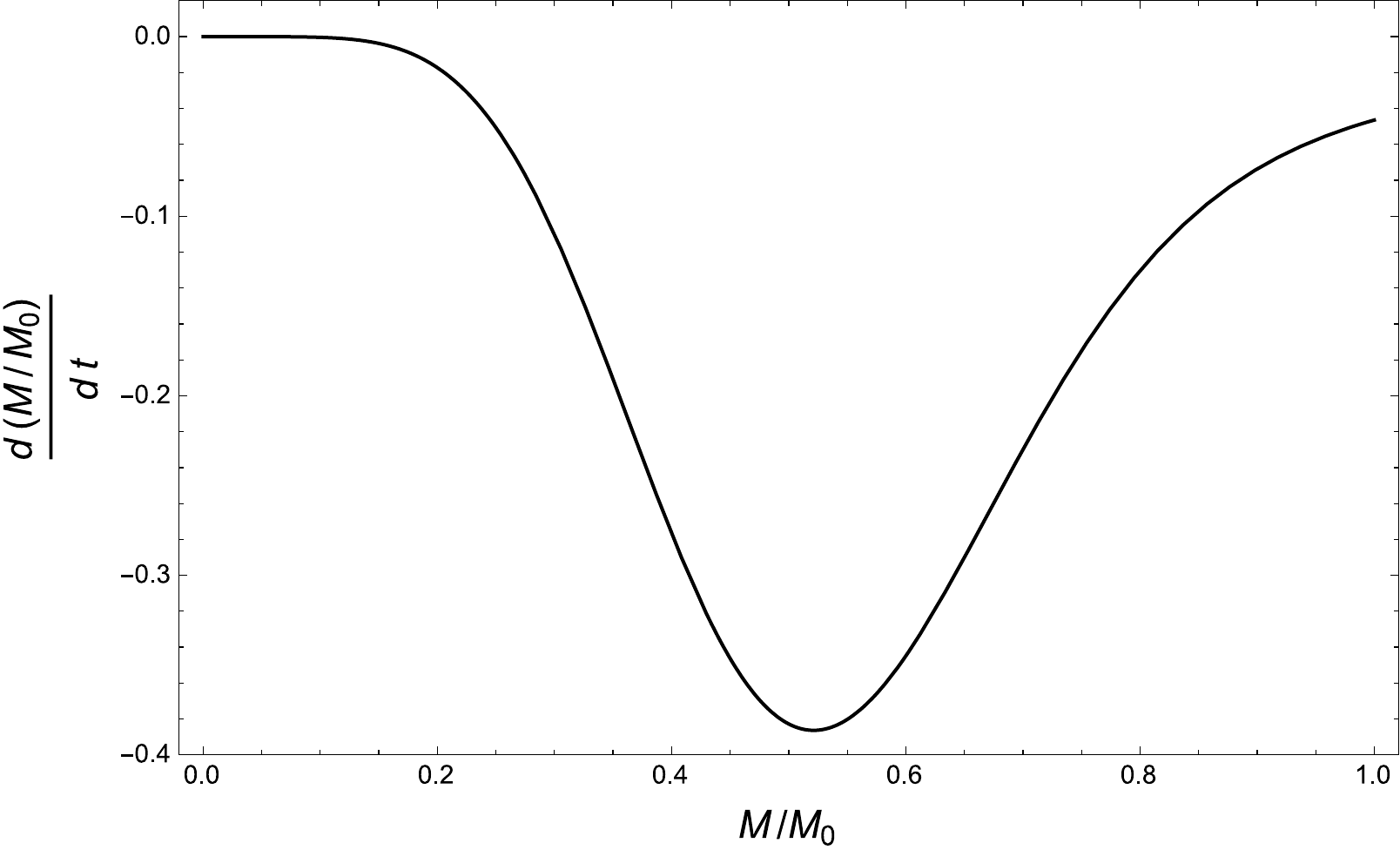}
\end{center}
\vspace{-0.2cm}
\caption{The reduced mass loss rate $dm/dt$ as a function of the reduced black hole mass $m$ within the micro-canonical ensemble assuming that the initial mass is $M_0=1$. $L=1$ in Planck units (left panel) and $L = \alpha M$ with $\alpha=4$ (right panel). We used $\sigma=\sigma_{\rm SB}$ for simplicity. 
\label{f-dmdt}}
\end{figure*}

\begin{figure*}[t]
\vspace{0.3cm}
\begin{center}
\includegraphics[type=pdf,ext=.pdf,read=.pdf,width=7cm]{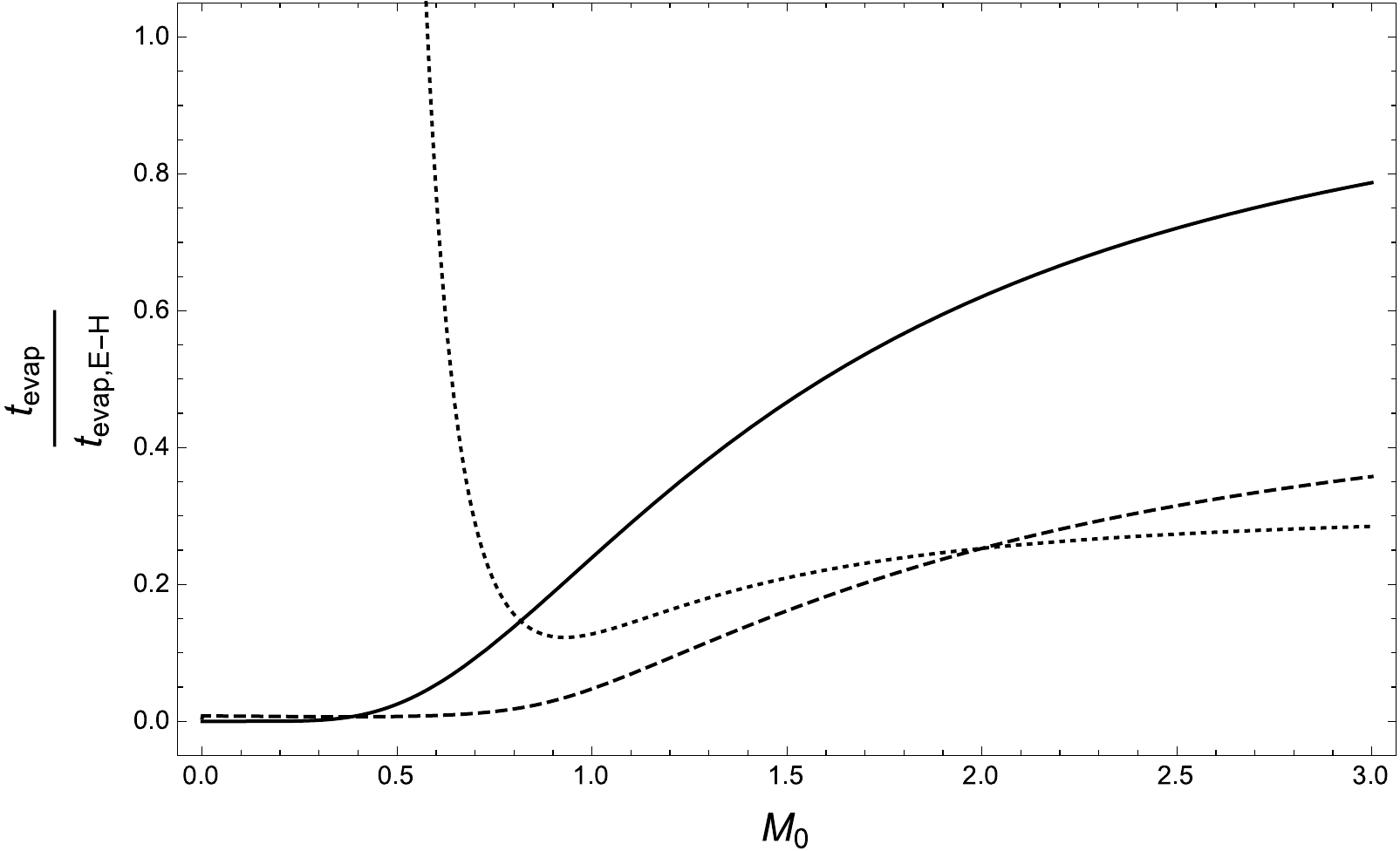}
\end{center}
\vspace{-0.2cm}
\caption{The evaporation time normalized to the one in Einstein's gravity $t_{\rm evap,\,E-H}$ as a function of the initial mass $M_0$ of the black hole (in Planck units). We show the following cases: $L=1$ in the canonical ensemble from formula \eqref{canLconst} (solid), $L=1$ in the micro-canonical from \eqref{tevapLconst} (dashed), and $L=0.5M_0$ in the micro-canonical ensemble from \eqref{tevapLconst} (dotted). \label{tevap}}
\end{figure*}

If we integrate both sides of Eq.~(\ref{eq-dmdt}), we find the evaporation time. In Einstein's gravity, $L = 0$, and the evaporation time is approximately
\be\label{eq-evap1}
t_{\rm evap,\,E-H} \approx \frac{256 \pi^3}{3 \sigma_0} M^3_0 \, ,
\label{tevapSch}
\ee
where $\sigma_0 = \sigma (M_0)$ and $M_0$ is the initial mass of the black hole. Since the black hole spends most of its life near its original mass, we can ignore the dependence on the mass $M$ in $\sigma$, simplifying the analysis. If $L\neq0$ is a constant, then we find
\be\label{eq-evap2}
t_{\rm evap} &=& \frac{2 \pi^3}{3 \sigma_0} 
\Bigg[ \frac{2 M_0}{\left(L^2+4 M_0^2\right)^3} 
\Big(4096 M_0^8 - 9216 L^2 M_0^6  
- 11088 L^4 M_0^4 - 3360 L^6 M_0^2 - 315 L^8 \Big) \nonumber\\ 
&& \hspace{1.5cm}+ 315 L^3 \arctan \left(\frac{2 M_0}{L}\right) \Bigg]       \, .
\label{canLconst}
\ee
If $L$ is very small ($L\ll M_0$), like the Planck length, the corrections in Eq.~(\ref{eq-evap2}) are negligible and we recover Eq.~(\ref{eq-evap1}). If $L = \alpha M$, for some constant $\alpha$, the evaporation time becomes
\be
t_{\rm evap} =  \left(1 + \frac{\alpha ^2}{4} \right)^{\!\! -4} 
\frac{256 \pi^3}{3 \sigma_0} M^3_0 \, .
\label{canLvar}
\ee
The case $L \propto M$ is motivated by the intriguing possibility that the length scale $L$ at which conformal symmetry becomes important depends on the actual mass $M$ of the object. It is useful to notice that the evaporation time $t_{\rm evap}$ can be written as proportional to the cube of the initial mass $M_0$ of the object and only depends on the ratio $\mu=L/M_0$. For the formula~\eqref{canLconst}, we find
\be
 t_{\rm evap} \! = \! \frac{\pi^3 M_0^3}{3 \sigma_0}
\Bigg[ \left(1+\frac{\mu^2}{4} \right)^{\!\! -3}
\Big(256  - 576 \mu^{2}- 693 \mu^{4}  
 - 210  \mu^6 - \frac{315}{16} \mu^{8} \Big) + 630 \mu^{3}   \arctan (2\mu^{-1}) \Bigg]  , 
 \nonumber 
\ee
while it was obviously already applied in the formula~\eqref{canLvar}. We notice that in all cases the luminosity $L_{\rm H}$ diverges in the last moment of evaporation $M=0$.

\subsection{Micro-canonical ensemble}

In the micro-canonical ensemble, the statistical system is kept in contact with a thermal reservoir, so the temperature $T$ is fixed. Employing the micro-canonical ensemble~\cite{casadio1,casadio2, hossenfelder}, we find more substantial differences with respect to the standard case in Einstein's gravity. We use the fact that in the micro-canonical ensemble the entropy is derived from the relation $\frac{dS}{dM}=\frac{1}{T}$ and we find that the entropy is given by the same formula as in the Schwarzschild case, namely $S(M)=4\pi M^2$. Moreover, the formula for evaporation rate in the micro-canonical ensemble is~\cite{hossenfelder}
\be
\frac{dM}{dt}=-{\cal E}A\,,
\ee
where ${\cal E}$ is the energy density of the radiation field surrounding the evaporating black hole. This reads in full generality
\be
{\cal E}=\frac{\sigma}{3}e^{-S(M)}\int_{0}^{M}e^{S(x)}(M-x)^{3}dx\,.
\ee
For the case of Einstein's  or conformal gravity this is explicitly
\be
{\cal E} = \frac{\sigma}{96\pi^2}e^{-4\pi M^2}\left(\vphantom{\frac{1}{2}}\pi M  \left(8 \pi  M^2-3\right)  \text{erfi}\left(2 \sqrt{\pi } M\right)\right. \left.
+12 \pi  M^2+e^{4 \pi  M^2} \left(1-4 \pi 
   M^2\right)-1\right), \nonumber\\
\ee
where the ${\rm erfi}(z)$ is an imaginary error function ${\rm erf}(iz)/i$, which takes real values for real arguments. We remind here that the black hole mass $M$ is measured in Planck units, hence $M$ in the above formula is dimensionless.

If $L$ is a constant, then the formula for the evolution of the black hole mass is
\be
\frac{dM}{dt} &=& -\frac{ \sigma\, e^{-4 \pi  M^2}}{6\pi\, M^6} 
\left(M^2+\frac{L^2}{4}\right)^4
\Big[ \pi M  \left(8 \pi  M^2-3\right)  \text{erfi}\left(2 \sqrt{\pi } M\right)+12 \pi  M^2
\nonumber\\ && \hspace{4.5cm}
+e^{4 \pi  M^2} \left(1-4 \pi 
   M^2\right)-1  \Big].
   \label{microLconst}
\ee
It is remarkable that the evaporation rate depends explicitly on the length scale $L$, and cannot be written as a function of only the reduced mass variable $m =  M/M_0$. The plot of $dm/dt$ as a function of the reduced black hole mass for a particular black hole with the initial mass $M_0=1$ and for the length scale $L=1$ in Planck units is shown on the left panel in Fig.~\ref{f-dmdt}. One sees that the luminosity still diverges at the final moment of the evaporation, when $M = 0$. 

In the other case, when $L = \alpha M$, one finds
\be
\frac{dM}{dt} &=& - \frac{\sigma M^2 e^{-4 \pi  M^2}}{6\pi} 
\left(1+ \frac{\alpha^2}{4}\right)^4   
\Big[\pi M \left(8 \pi  M^2-3\right) \,
\text{erfi}\left(2 \sqrt{\pi } M\right)
+12 \pi  M^2 \nonumber\\
&& \hspace{4.8cm} +e^{4 \pi  M^2} \left(1-4 \pi  M^2\right)-1\Big] \, .
\ee
Now $dM/dt$ goes to zero for $M \rar 0$, as it is in the evaporation of black holes in the micro-canonical ensemble in Einstein's gravity. The plot of $dm/dt$ as a function of the reduced  black hole mass $m$ and when $L=\alpha M $ is shown in the right panel in Fig.~\ref{f-dmdt} for a particular black hole. The values chosen there are $M_0=1$, $L=1$ in Planck units and $\alpha=4$. It is worth noticing that the evaporation rates in both ensembles coincide for large values of the mass of the black hole (in the thermodynamical limit $M_0\gg1$) and this could be in principle observed for $m\approx1$, $m<1$ so on the left extreme of plots.

The explicit expressions for the evaporation time in the micro-canonical ensemble can only be obtained numerically by evaluating the following integrals
\be
\bar{t}_{\rm evap}=\!\int_0^{M_0}\! \frac{6\pi\,e^{4 \pi  M^2} dM}{\sigma  M^2 \left(\pi M \left(8 \pi 
   M^2-3\right) \text{erfi}\left(2 \sqrt{\pi } M\right)+12 \pi 
   M^2+e^{4 \pi  M^2} \left(1-4 \pi  M^2\right)-1\right)} \nonumber\\
\ee
in the standard Schwarzschild case,
\be
&& t_{\rm evap}(M_0,L)=  \label{tevapLconst}\\
&& \!\int_0^{M_0}\! \frac{{6\pi\, M^6}  e^{4 \pi  M^2} dM}{\sigma  \left(M^2+L^2/4\right){}^4 \left(\pi M \left(8 \pi 
   M^2-3\right) \text{erfi}\left(2 \sqrt{\pi } M\right)+12 \pi 
   M^2+e^{4 \pi  M^2} \left(1-4 \pi  M^2\right)-1\right)}
   \nonumber
\ee
in conformally modified Schwarzschild case with $L={\rm const}$, and
\be
&& t_{\rm evap} = \nonumber \\
&&  \!\int_0^{M_0}\!\!\!\! \frac{6\pi \, e^{4 \pi  M^2} dM}{\sigma  \left(1+ \frac{\alpha^2}{4}\right)^4 M^2 \left(\pi M \left(8 \pi 
   M^2-3\right) \text{erfi}\left(2 \sqrt{\pi } M\right)+12 \pi 
   M^2+e^{4 \pi  M^2} \left(1-4 \pi  M^2\right)-1\right)} \nonumber\\
   &&= \bar{t}_{\rm evap}\left(1+ \frac{\alpha^2}{4}\right)^{\!\! -4}
\ee
in the case $L=\alpha M$. However, we notice that the first and the third integral are not convergent due to a highly suppressed emission rate near $M=0$. For the first integral (standard Schwarzschild case), the mass loss rate ${dM}/{dt}$ behaves like $4/3\pi\sigma M^6$ and hence the evaporation time is formally infinite, because the rate ${dM}/{dt}$ decreases to zero too fast. In the case $L=\alpha M$, we fine the same evaporation time $\bar{t}_{\rm evap}$ with the additional multiplicative factor 
$\left(1+ {\alpha^2}/{4}\right)^{4}$. 
Only in the second case the integral is convergent near $M=0$, because the integrand goes there like 
$192 M^2/\pi  L^8 \sigma$, 
so the evaporation rate is still infinite at this last moment. This, on the other hand, ensures that the black hole lifetime is finite. However, in this case the evaporation time depends both on the initial mass $M_0$ and the length scale $L$.

The analysis of the evaporation rate at the last moment $M=0$ can be repeated with the results obtained in the canonical ensemble case. We find that in all three cases it diverges near the final moment of evaporation. In the original Schwarzschild case and in the case with $L=\alpha M$, the rate scales, respectively, like 
\be
\frac{\sigma}{256\, \pi ^3 M^2} \quad {\rm and} \quad  \frac{\sigma\left(1+\alpha ^2/4\right)^4}{256\, \pi ^3 M^2}. 
\ee
When $L={\rm const}$, we find that the rate diverges even stronger, because 
\be
\frac{dM}{dt}=\frac{\sigma L^8}{2^{16}\, \pi ^3 M^{10}}. 
\ee
Again this ensures finite black hole lifetime.

In all cases considered here, black holes completely evaporate without leaving any remnant. In some cases, the evaporation time is finite. In the micro-canonical ensemble, the evaporation time may be finite if $L={\rm const}$. In Fig.~\ref{tevap} we show the plots of the evaporation times normalized to the one in the standard Schwarzschild case \eqref{tevapSch}. Contrary to the case of Einstein's gravity, the lifetime derived in the micro-canonical ensemble is smaller than the corresponding one in the canonical ensemble. This is understandable, because the former ensemble tries to capture some effects, which could be traced back to the back-reaction and the time-dependence of the background spacetime of the black hole metric. Only in the micro-canonical ensemble and for $L=0$ or $L=\alpha M$ cases the mass loss rate is bounded at any time during the evaporation process, and this is physical. In the cases in which it diverges, the approximation used for deriving the Hawking evaporation rate is not valid. After all, the Hawking evaporation is a dynamical and non-stationary process.  We observe from Fig.~\ref{tevap} that in conformal gravity the evaporation time is always shorter in comparison with the same situation in Einstein's gravity. This can have important consequences for the astrophysics of primordial black holes.

\section{Penrose diagram \label{s-4}}

The Penrose diagram of the static black hole solution~(\ref{eq-k}) was discussed in Ref.~\cite{noi}. It is like the Penrose diagram of the maximally extended Schwarzschild-Kruskal spacetime, with the only difference that the space-like hypersurface $r = 0$ is now regular. This is possible because any observer needs an infinite time to reach the point $r=0$ and the spacetime turns out to be geodesically complete. In this section we want to discuss the Penrose diagram of the spacetime describing the gravitational collapse and the evaporation process. Note that the conformal diagrams of a static black hole and of a black hole formed by gravitational collapse are completely different, even in standard Einstein's gravity.

\begin{figure}[t]
\vspace{0.3cm}
\begin{center}
\includegraphics[type=pdf,ext=.pdf,read=.pdf,width=7cm]{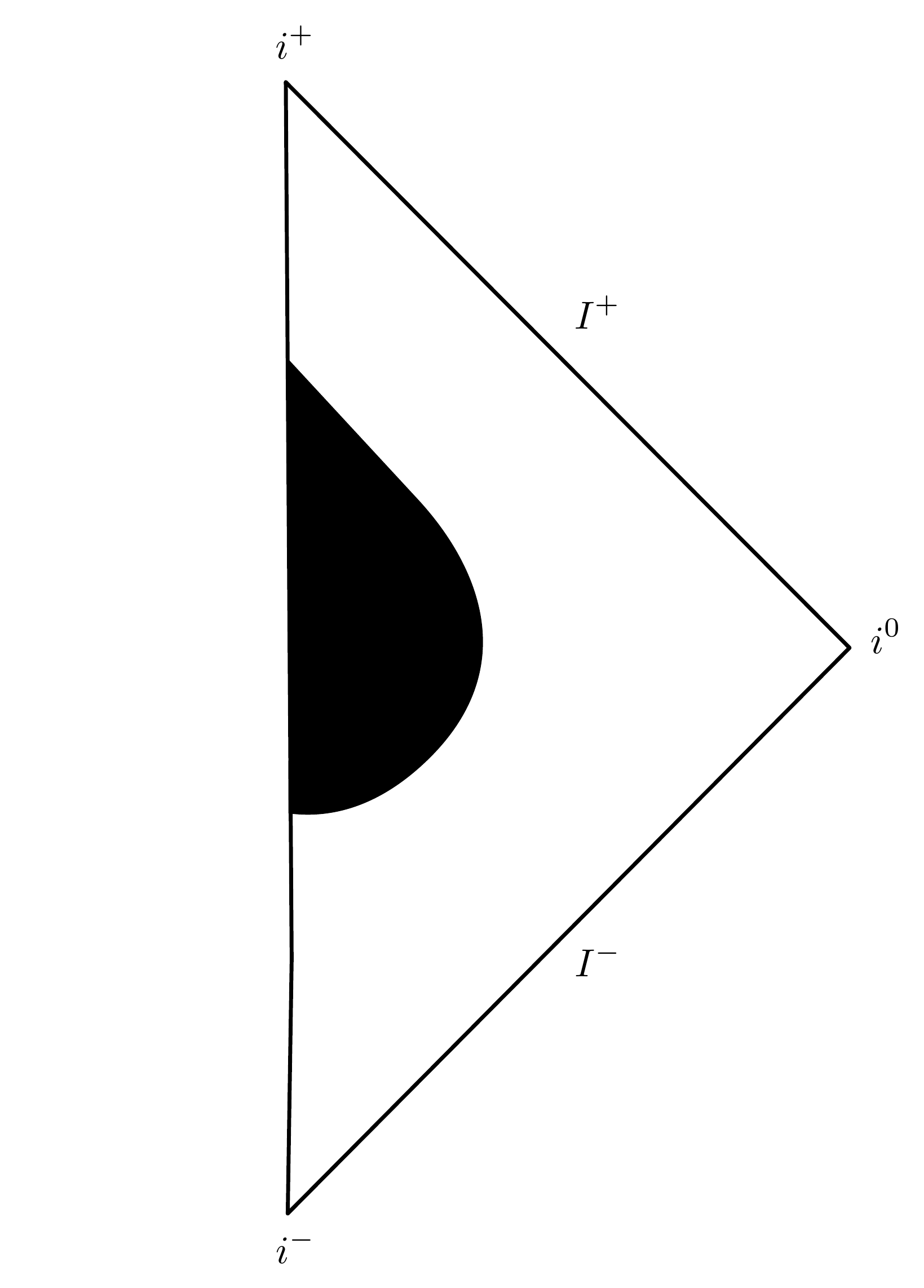}
\end{center}
\vspace{-0.2cm}
\caption{Penrose diagram for the formation and evaporation of a black hole in conformal gravity. See the text for more details. \label{f-penrose}}
\end{figure}

The evaporation process discussed in the previous section fits in the class of ``complete evaporation scenarios'' of Ref.~\cite{smolin}. Its Penrose diagram, joined to the collapse described in Section~\ref{s-2}, is sketched in Fig.~\ref{f-penrose}. The black area represents the region of the spacetime inside a trapped surface. Strictly speaking, there is no black hole here, because there is no event horizon but only a trapped surface that lasts for a finite amount of time. Note that the diagram in Fig.~\ref{f-penrose} follows from two simple considerations: $i)$ the absence of the singularity at $r=0$, and $ii)$ the fact that the evaporation time is finite. See also Ref.~\cite{b3} for the construction of this diagram. If $L\propto M$, the mass $M$ goes to zero smoothly in a finite amount time and all curvature invariants go to zero too. If $L$ is a constant independent of $M$, the spacetime is not flat, but it is conformal to the flat spacetime and the Penrose diagram after evaporation coincides with the Penrose diagram of the Minkowski spacetime.

The point at $r=0$ is always regular. There is no singularity of curvature there, neither of geodesics. The singularity of the geodesics is not solved by extending the spacetime to negative $r$ (attaching another universe with a gate to it at $r=0$) nor by introducing an inner horizon, but by the fact that particles cannot reach the center of the black hole at $r=0$ in a finite proper time (massive particles) or finite value of the affine parameter (massless particles). The Penrose diagram is similar to those of the models studied in Refs.~\cite{f1,f2,b1,b2,b3,b4,b5,b6}. The trapped surface first behaves as the horizon of the black hole, and later as the horizon of a white hole. After the evaporation of the black hole, there is no remnant, and therefore the final spacetime looks like the Minkowski flat spacetime (in conformal gravity, it is correct only to say that the final spacetime is conformally flat). However, the Penrose diagram is not symmetric with respect to the horizontal axis $t=0$ as in the case of the ``black supernova'' scenario discussed in Ref.~\cite{b3} (see Figs.~1 and 2 in~\cite{b3}). Here in conformal gravity, we have (classically, i.e. for $\hbar \rar 0$) the formation of a true black hole (Fig.~4 in~\cite{b3}). The evaporation is due to the Hawking radiation ($\hbar \neq 0$), not to the bounce of the collapsing shell \cite{f0,f1,f2,f3,f4,f5,f6,b1,b2,b3, b4}. For a distant observer, the timescales of collapse and of evaporation are very different: the former is short and the latter is very long. This justifies the fact we use the time-dependent conformally transformed Vaidya metric to describe the dynamical collapse of radiation, and the time-independent conformally-transformed Schwarzschild metric to describe the evaporation process.

Last, we note that our Penrose diagram should not be affected by possible back reaction effects. Indeed, we expect that in a ultraviolet completion of the theory (\ref{confEgrav}) (see for example \cite{p1}) conformal symmetry is restored at high energy. Therefore, only a trivial S-Matrix ($S=1$) is compatible with the Coleman-Mandula theorem. In such a case, there are no interactions at high energies and thus there is no back reaction to take into account.

\section{Concluding remarks \label{s-5}}

Conformal gravity is a viable scenario to solve the problem of spacetime singularities that appears in Einstein's gravity. In this paper, we have studied the formation and the evaporation of a singularity-free non-rotating black hole and guessed the corresponding Penrose diagram for this physical process. For the formation of the black hole, we have considered the collapse of a thin shell of radiation, which can be trivially obtained from the solution in Einstein's gravity by applying a suitable conformal transformation. We have used the Vaidya metric ansatz, which is an exact solution in both Einstein's and conformal gravity. This solution is consistently sourced by the conformal matter stress-energy tensor.

For the evaporation process, we have studied the corresponding black hole thermodynamics. We have employed the ansatz of the static conformally transformed Schwarzschild spacetime. The black hole temperature is the same as in Einstein's gravity, which had to be expected restricting our attention to massless particles only (required by conformal symmetry). However, the surface area is different. Depending on the interpretation of the new scale $L$, the evaporation time may deviate from the standard prediction in Einstein's gravity, but we never find a significant difference, in the sense that the evaporation time computed in the canonical ensemble is still $t_{\rm evap} \propto M^3_0$, where $M_0$ is the initial mass of the black hole (compare also with Fig.~\ref{tevap}).  We also studied the evaporation process in the micro-canonical statistical ensemble finding stricter bounds on the evaporation rates.

The Penrose diagram of the formation and complete evaporation of a black hole in conformal gravity is sketched in Fig.~\ref{f-penrose}. Since particles can never reach the centre at $r=0$ in a finite amount of time, there is no other universe and the Penrose diagram are similar to those of the bouncing scenarios of Refs.~\cite{f0,f1,f2,f3,f4,f5,f6,b1,b2,b3}.

%%%%%%%%%%%%%%%%%%%%%%%%%%%%%%%

\begin{acknowledgments}
C.B. acknowledges support from the NSFC (grants 11305038 and U1531117), the Thousand Young Talents Program, and the Alexander von Humboldt Foundation. S.P. thanks the Department of Physics at Fudan University for hospitality during his visit.
\end{acknowledgments}

%%%%%%%%%%%%%%%%%%%%%%%%%%%%%%%

\appendix

\section{Scalar invariants of Vaidya spacetime \label{s-a}}

Scalar invariants in the conformally modified Vaidya spacetime are everywhere regular and do not diverge at $r = 0$ for $L \neq 0$. To prove this assertion, we list some scalar invariant functions (containing up to four derivatives of the metric). The Kretschmann scalar is:
\be
R_{\mu\nu\rho\sigma} R^{\mu\nu\rho\sigma} &=&
\frac{48r^{10}M_{0}\Theta(v-v_{0})}{\left(L^{2}+r^{2}\right)^{12}}
\big[3L^{8}(59M_{0}-48r)+4L^{6}r^{2}(9M_{0}+8r)
\nonumber\\ &&
\hspace{3.5cm} +2L^{4}r^{4}(91M_{0}-72r)+4L^{2}M_{0}r^{6}+M_{0}r^{8}\big] \nonumber\\&&
  +\frac{64r^{12}\left(23L^{8}-14L^{6}r^{2}+23L^{4}r^{4}\right)}{\left(L^{2}+r^{2}\right)^{12}}
    \, .
\ee
The square of the Ricci tensor is:
\be
 {\bf Ric}^2 &=& \frac{64 L^4 r^{10} }{\left(L^2+r^2\right)^{12}}
  \big[6 M_{0} \Theta (v-v_{0}) \big(3 L^4 (9 M_{0}-7 r)+2 L^2 r^2 (9 M_{0}-4 r)
  \nonumber\\ &&
 \hspace{2.2cm} +r^4 (15 M_{0}-11 r)\big)+25 L^4 r^2+2
   L^2 r^4+13 r^6\big] \, , 
\ee
where ${\bf Ric}^2 \equiv R_{\mu\nu} R^{\mu\nu}$. 
Finally, the scalar curvature is:
\be
R = -\frac{24 L^2 r^5}{\left(L^2+r^2\right)^6} 
\Big[-4 M_{0} \left(2 L^2+r^2\right) \Theta (v- v_{0})+3 L^2 r+r^3\Big]        \, .
\ee
The modified Vaidya metric has thus no curvature singularities. As in the case of the modified Schwarzschild metric~\cite{p1,noi}, one can also check that this spacetime is \emph{not} geodesically incomplete at $r=0$ because the origin $r=0$ is reached with an infinite value of the geodesic affine parameter. For example, massive particles never reach (or come out from) the point $r=0$ in a finite amount of their proper time.

\end{document}